\documentstyle[preprint,aps]{revtex}
\begin{document}
\vskip2cm
\title
{ Enhancement of top production in $e^+ e^- \rightarrow t\bar t$}
\author{
Lali Chatterjee$^{1,2}$ and Cheuk-Yin Wong$^1$}
\address{
${}^1$Oak Ridge National Laboratory, Oak Ridge, TN 37831
}
\address {\rm and }
\address{ ${}^2$Jadavpur University, Calcutta 700032, India.}
\maketitle
\vspace{2cm}
\begin{abstract}

\noindent
We investigate $e^+ e^- \rightarrow t\bar t$ by taking into account
wave function distortion effects in the near-threshold region and
interpolating the correction factor to the high-energy perturbative
QCD region.  The strong color attraction between the color singlet
$t$ and $\bar t$ enhances the cross section over the tree level results
near threshold.  The rise of the cross section near the threshold is
smoothed by the effects of the large top width and below-threshold
resonances.  The reliability of the prediction is well assessed by
comparing with experimental $R$ ratios for lighter quark production at
lower energies.  The cross sections obtained here using the reported
values of the top mass, can be used directly for New Linear Collider
(NLC) projections.

\end{abstract}
\pacs{ PACS number: 13.65.+i }

\narrowtext
\newpage
\section{ Introduction}

Recent reports of the observation of the top quark, with a mass of
$174$ GeV
\cite{Abe94}, provoke improved theoretical studies of the top flavor.
  The planned and proposed experimental focus for the next decade to
develop and understand top physics and top characteristics add impetus
to theoretical investigations
\cite{Che94,Eic94b,Lan94,Hil94,Ber94,Lae94}.
  The super-heavy value of the top mass also opens up new vistas of
electroweak physics and possibilities for probing for effects beyond
the Standard Model.  Many special features of the top have been
studied recently\cite{Fad90,Str91,Kwo91,Sum93,Ynd94}.  Being heavier
than the W, the top is extremely susceptible to weak decay by the
$t\rightarrow W^+ b$ channel. This large width distinguishes it from
its doublet partner $b$, as well as the lower generation quarks and
will lead to the disappearance of the sharp resonance structure
corresponding to below-threshold resonances, as one transits from the
$c$ and $b$ to the $t$ flavor.  Yet another feature, special to the
top sector is the reliability of perturbative QCD to probe its bound
and continuum states as the non-perturbative regime does not
contribute effectively at the very short distances characterized by
$1/m_t$ for top flavor production.

 Careful analyses of higher-order QCD corrections for production of
heavy quarks with a finite mass have been carried out
\cite{Bee89,Nas89a,Nas89b} which indicate that the next-to-leading
order corrections are very large when the relative velocity between
the top quark and the top antiquark is small. For such massive
fermions near the threshold, the distortion of the outgoing relative
wavefunctions between the top quark and the top antiquark, due to the
strong color field connecting them, would be a dominant correction to
the $t\bar t$ production processes.  This can be understood in terms
of `Coulomb gluons' analogous to those acting between $c$ and $\bar c$
in the $J/\psi$ systems and between $b$ and $\bar b$ in $\Upsilon$.
The importance of such Coulomb effects for heavy quark production
processes has been recognized earlier in the study of heavy quark
productions in bound and continuum states
\cite{App75,Bar80,Gus88,Fad88,Fad90}.

In the present work we expand on these aspects and predict the rates
for $t\bar t$ production by virtual $\gamma$ and $Z$ exchange in $e^+
e^-$ collider experiments, according to $ e^+ e^- \rightarrow \gamma^*
(Z^*) \rightarrow Q\bar Q$, including the important near-threshold
region. The two loop modified running coupling constant $\alpha_s
(q^2)$ in the five flavor theory is used and the threshold QCD
corrections are folded into the high energy results by an
interpolation formula. Effects of top width and below-threshold
resonances are included. Corrections for running of the QED coupling
have been incorporated.  The cross section obtained here using the
latest top mass, can be used for future $e^+e^-$ collider projections.
Results are extended to the charm and beauty flavors and applied to
the familiar $R$ parameter for comparison with experiment.

\section{ Tree level cross section}

The differential cross section for $e^+e^- \rightarrow Q \bar Q$ above
threshold at a center-of-mass energy $\sqrt{s}$ is given at the tree
level by Eqs.\ (24.11)-(24.19) of Ref.\ \cite{Rev94},
\begin{eqnarray}
\label{eq:1}
 { d \sigma (s) \over d z } =
{ 3 \pi \alpha^2 \over 4 s }
\biggl ( {e_Q \over e } \biggr )^2
 {\beta D} ,,
\end{eqnarray}
where $z=\cos \theta$, and $m_Q$ and $e_Q$ are the quark mass and
charge, and we have included the three different colors of the
quark-antiquark pair.  In Eq.\ (\ref{eq:1}), the quantity $\beta$ is
$$\beta=
\sqrt{1- {4 m_Q^2 \over s}}\,,
$$
the quantity  $D$ is
\begin {eqnarray}
D=
{e_Q^2 A -2 e_Q \chi_1 [V V_f A -2 a_f \beta z]+ {\chi_1}^2 B }\,,
\end {eqnarray}
where
$$
A=
{1 +z^2 +(1 - \beta^2) ( 1-z^2)},
$$
$$
B=
{ V_f^2 (1+V^2) A
+\beta^2 a_f^2 (1+V^2) [1 +z^2] -8 \beta V V_f a_f z},
$$
$$ \chi_1 =
{ x \over {16 \sin^2 \theta_w \cos^2 \theta_w}}, $$
$$
V=
{-1+4 \sin^2 \theta_w }\,,
$$
$$
V_f=
{2 T_{3f}- 4 e_Q \sin^2 \theta_w}\,,
$$
$a_f =2 T_{3f} =1$ for the top, and
 $\theta_w$ is the Weinberg angle.
The quantity $x$ contains the $Z$ mass $m_Z$ as
$$
x =
 { s \over {s -{m_Z^2 }}}.
$$
Integrating over the angle, as we are not probing the asymmetry but the
 top production, we obtain
\begin{eqnarray}
\label{eq:sigs}
\sigma(s) =
{4 \pi \alpha^2  \over s} \biggl ( {e_Q \over e } \biggr )^2
H \beta
\end{eqnarray}
where
\begin{eqnarray}
H=
\biggl ( {1 + {1 -\beta^2\over 2}}\biggr )
\biggl \{ { e_Q^2 -2 e_Q \chi_1 V V_f +\chi_1^2 V_f^2(1+V^2)}\biggr \}
    +\beta^2 a_f^2 (1+V^2)\,.
\end {eqnarray}
  The above expressions are obtained using plane waves for all the
fermions.  It is however well recognized that the lowest order
diagrams are inadequate to describe the quark sector, because of the
large value of the QCD coupling constant $\alpha_s$.  Even for light
quarks, the next-to-leading order radiative corrections for such
production vertices predict an increase in the lowest
order values, according to the correction factor $(1 + {\alpha_{s}}/
\pi)$, where $\alpha_{s}$ is the strong coupling constant \cite
{Fie89}.  These corrections, obtained painstakingly as the residual
quantities after cancellation of singularities, unfortunately require
the massless or chiral limit for their applicability.  It is easy to
see that the massless limit is not a desirable {\it modus operandi}
when dealing the top flavor near the threshold, with its mass of $174$
GeV.

    The large mass of the top quark assures the importance of the part
of the vertex correction due to the exchange of longitudinal gluons
between the top quark and the top antiquark at the $\gamma^* t \bar
t$, or $Z^* t \bar t$ vertex, near the $t\bar t$ threshold at around
$350$ GeV. This exchange can be represented by a potential that acts
between the $t$ and $\bar t$.

\section{ Vertex correction factor}

If we consider direct materialization of a quark-antiquark pair by
decay of an intermediate virtual photon for instance, the process is
characterized by a region of very small relative $Q$-$\bar Q$
separations, at a $Q$-$\bar Q$ distance of $\sim \alpha/\sqrt{s}$.  The
effective part of the $Q$-$\bar Q$ interaction in this region of small
relative separation associated with the production process is the
inverse-$r$, Coulomb-like term. The linear part of the potential, that
serves to effect the binding and impose confinement, controls the
large $r$ behavior \cite{Eic80,Eic94a} and needs not be considered in
the collapsed spatial zone comprising the annihilation or production
vertex.  This is further augmented by the short lifetime of the $t$
and $\bar t$ that prohibits their moving away beyond the perturbative
range.

Upon expressing the one-gluon exchange potential between the quark and
the antiquark in the $1/r$ Coulomb-like form, one can obtain exact
solution for the wavefunction \cite{Akh65} and extract the $r=0$ part
to correct the plane waves used in the Born diagram for their
distortion in the color field.  This procedure corrects for the
distortion effect to all orders, and the resulting corrections can
extend the applicability of PQCD for this process to the near-threshold
region. We have discussed these correction factors for the lighter
quark annihilation and production in a separate work\cite {Cha94}.
Though we restrict the main analysis to top flavor, the results can be
easily extended to lighter flavors, and are seen below to be
compatible with charm and beauty production near their relevant
thresholds.

 We use a suitable short distance color potential of the Coulomb type,
$$ V(r) =  {- \alpha_{\rm eff} \over r}, $$
where
$$
\alpha_{\rm eff} =
C_f \alpha_s(q^2),
$$
$C_f$ is the appropriate color factor,
and $\alpha_s$
is the two-loop modified running QCD coupling constant and can be calculated
from \cite{Rev92}
\begin {eqnarray}
\label{eq:coup}
\alpha_s(q^2)=
{4 \pi \over (11-2n_f/3)  \ln (q^2/\Lambda^2)} \biggl (1-{2 (51-19n_f/3)
\over (11-2n_f/3)^2}
{\ln[\ln(q^2/\Lambda^2)]\over \ln(q^2/\Lambda^2)} \biggr )\,,
\end{eqnarray}
where $q^2$ is identified as $s$ and $\Lambda$ is taken to be
$\Lambda^{(5)}=195$ MeV for top production.

For outgoing final states relevant to a produced
$Q\bar Q$ pair, the wave function of the quark $Q$
in the field  of the antiquark $\bar Q$
is given by\cite{Akh65}
\begin{eqnarray}
\psi = N e^{i {\bf p} \cdot {\bf r}}
(1- {i \over 2E} \bbox{\alpha} \cdot \nabla )
u F(-i \xi, 1, -i(pr +\bbox{p}\cdot \bbox{r})),
\end{eqnarray}
where $u$ is the spinor for the free quark,
$F$ is the confluent hypergeometrical function
and $N$ is the normalization constant given by
$$
|N|^2= {2 \pi \xi \over 1 - e^{-2 \pi \xi} }\,,
$$
where
$$
\xi= \alpha_{\rm eff} / v,
$$
and $v$ is the relative velocity between the quark and the antiquark
given by \cite{Cha94}
\begin{eqnarray}
\label{eq:vvv}
v=
{\sqrt{1 - 4  m_Q^2/s} \over 1 - 2 m_Q^2/s}.
\end{eqnarray}

The square of the wavefunction taken at contact can be obtained as a
generalization of the familiar `Gamow-Sommerfeld factor' \cite {Gam28}
as
\begin{eqnarray}
\label{eq:kv}
K^{(f)}
=|\psi(0)|^2 =  { 2 \pi \xi \over 1 - e^{-2 \pi \xi} }
(1 + \alpha_{\rm eff}^2),
\end{eqnarray}
where  spins of both quarks have been averaged over and
the superscript $(f)$ is to indicate that this is the correction
factor for quark-antiquark production in the final state \cite{Cha94}.

 In order to obtain a generalized correction factor, that is valid not
only when $Q$ and $\bar Q$ have low relative velocities, but also have
large relative velocities, we use the interpolation technique
suggested by Schwinger\cite{Sch73}.  In the massless quark limit, the
next-to-leading order QCD corrections to production of $q\bar q$ at
high energies is given by the correction factor $K^{(f)} =(1
+{\alpha_s/ \pi})$ \cite {Fie89}. The transition of the QCD
corrections from the correction factor $ K^{(f)}$ of Eq.\
(\ref{eq:kv}) at low relative velocities, to the $( 1 +{\alpha_s /
\pi} )$ behavior at relativistic velocities can be accommodated with
the introduction of an interpolative function $f^{(f)}(v)$, where
\begin{eqnarray}
f^{(f)}(v)=
{\alpha_{\rm eff} \biggl [ {1 \over v}
+ v \biggl ( -1 + {3 \over 4 \pi^2} \biggr ) \biggr]}.
\end{eqnarray}
The generalized QCD correction factor, applicable at all relative velocities
  can be
written as
\begin{eqnarray}
\label{eq:kvall}
K^{(f)} =
{ 2 \pi f^{(f)}(v) \over { 1 - exp[-2 \pi f^{(f)} (v)]} }  (1+ \alpha_{\rm
eff}^2).
\end{eqnarray}

Incorporating the distortion correction, equation (\ref{eq:sigs}) for
$t\bar t$ formation by the electroweak channels becomes, at energies
above 2$m_Q$,
\begin{eqnarray}
\label{eq:sig}
\sigma (s) = K^{(f)}  {4 \pi   \alpha^2  \over s}
\biggl ( {e_Q \over e } \biggr )^2
\sqrt{1-{4 {m_Q}^2 \over s}}
H
\end{eqnarray}
For virtual photon or $Z$ mediated processes, the quark-antiquark pair
must be in color-singlet states.  The color factor $C_f$ has the value
$4/3$ and the one-gluon exchange potential is attractive between the
quark and antiquark.  Due to this attractive nature of the
color-singlet interaction, the correction factor (\ref{eq:kvall})
enhances top production over the tree level expectations.  As in
earlier analyses \cite{Fad90},\cite{App75}-\cite{Fad88}, there is a
large enhancement of top flavor production near threshold due to the
wavefunction distortion by this attractive color field.

It may be noted that in the large $\xi$ limit, the correction factor
in Eq.\ (\ref{eq:sig}) takes the form $ \pi \xi= \pi C_f \alpha_s/v$.
As $v \rightarrow 0$, corrected $\sigma(s)$ of Eq.\ (\ref{eq:sig})
does not diverge to infinity, because the $\sqrt{1-4m_Q^2/s}$ term of
(\ref{eq:sig}) cancels with the corresponding term in the velocity in
Eq.\ (\ref{eq:vvv}).  This cancellation of the Coulomb singularity by
the phase space factor has been discussed in the context of $Q\bar Q$
formation by $q\bar q$ annihilation in hadronic collisions
\cite{Bee89}.  In the absence of the vertex correction, the top
production rate would plunge to zero at threshold, whereas with the
correct inclusion of distortion effects, the cross section attains a
relatively constant value. The introduction of the bound toponium
states and the finite top width modifies this constancy as one
approaches the threshold and softens considerably the cut-off at
the threshold.

\section{  Inclusion of the top decay width and below-threshold bound states}

    The large weak decay rate of the top and the contribution of
below-threshold bound states should be included for correct prediction of
near-threshold top production. In the Standard Model, the decay rate
for the top by the dominant $ t \rightarrow W^+ b $ mode can be written
as \cite{Big86}
\begin {eqnarray}
\Gamma_t =
{ G_f m_t^3 \over 8 \pi\sqrt{2}} |V(tb)|^2 {2 k\over m_t} L\,,
\end {eqnarray}
where
$$
L=
[1-(m_b/m_t)^2]^2 +{ [1 + (m_b/m_t)^2] [(m_w/m_t)^2] } - 2 (m_w/m_t)^4\,,
$$
$$
k=
{ \biggl ( [m_t^2-(m_w+m_b)^2]^{1/2} [m_t^2-(m_w -m_b)^2]^{1/2}\biggr )
  \over 2 m_t}\,,
$$
and $V(tb)$ is the Standard Model mixing parameter which can be taken to be
unity.

    The effects of the top width can be introduced by including
the contribution from all other energies to
the cross section at the energy under consideration.
The cross section at a given energy $\sqrt{s}$ should then be written as
\begin{eqnarray}
\sigma(s)=
\sum_{i} \sigma_i (E_i)  J_i (s,E_i) \,,
\end{eqnarray}
where $\sigma_i(E_i)$ is the total cross section for state $E_i$
and $J_i(s,E_i)$ is the normalized Breit-Wigner distribution
$$
J_i (s, E_i)=
{1\over \pi}  ~{ {\Gamma/ 2} \over{ [ (\sqrt{s}-2m_t-E_i)^2 +(\Gamma/2)^2]} }
\,.
$$
The summation runs over all state energies $E_i$ and the width
$\Gamma$ refers to the decay of the combined $t\bar t$ system and is
given by $ \Gamma = 2 \Gamma_t$.  The cross section can be devolved
into a summation $\sigma_b(s)$ over all discrete bound states and an
integration $\sigma_c(s)$ over all continuum states as
\begin{eqnarray}
\sigma(s)=
\sigma_b (s)  + \sigma_c(s)\,.
\end{eqnarray}
Integrating over the continuum, the open or continuum cross section, including
the top width effects can be written as
\begin{eqnarray}
\sigma_c(s)=
 K^{(f)} {4 \pi \alpha^2 \over s}
\biggl ( {e_Q \over e } \biggr )^2
\beta   H J_c(s)\,,
\end{eqnarray}
where
$$
J_c(s)=
{1\over \pi } ~\biggl \{ {\pi\over 2} + \tan^{-1}
\biggl ({\sqrt{s}-2m_t \over \Gamma} \biggr )
 \biggr \} \,.
$$
In this case the width softens the sharp step-function type onset of continuum
production into a smooth leakage into the  region below the threshold.

  The case of the bound states requires the input of the bound state
energies and wavefunctions.  It is expected that the attractive color
field between the produced $t\bar t$ will form bound toponium states
like the charmonium and bottonium systems with similar spectroscopic
properties. The $t\bar t$ system differs from the $c\bar c$ and $b\bar
b$ in two important respects. Due to the large virtuality of the
process for the pair with top flavor and the correspondingly large QCD
scale, characterized by $1/m_t$, the QCD perturbative coupling is
considerably smaller than that for the lighter $q\bar q$ pairs and PQCD is
justified. Secondly, the large top mass and resulting large decay
probability into the $W$ and $b$ is a characteristic not shared by the
pairs with $b$ or $c$ flavor.

The large magnitude of the momentum transfer allows us to describe the
lowlying toponium bound states as those in the attractive perturbative
Coulomb-like inter-quark color field as nonperturbative effects are
not manifested at the short distances applicable to $t\bar t$
materialization. The effects of the gluon condensate have\ been
studied by \cite{Ynd94,Fad90} and shown to be small. The work in
\cite{Ynd94} further establishes the validity of the Coulomb color
potential for the toponia states.

Bound states in the $t\bar t$ color field can be described simply by
Coulomb-like wavefunctions and energy levels.  Using the same
perturbative Coulomb-like potential as for the continuum, we can write
for the energy of the $n$th bound state
$$
E_n=
 - {{\mu {\alpha_{\rm eff}}}^2 \over {2 n^2}}\,,
$$
where $\mu $ is the reduced mass of the quarks.
The wavefunction at $t\bar t$ contact can be written  as \cite{Kwo91}

$$
| \psi_n (0) |^2
= {1\over \pi} \biggl ({\mu \alpha_{\rm eff} \over n} \biggr )^3 \,.
$$
The total cross section for below-threshold
resonance toponium production can then be written as
\cite{Kwo91}
$$
 \sigma_n =
{ 4 \pi \alpha^2 \over s}\biggl ( {e_Q \over e} \biggr )^2
{}~ {3 \pi m_t \alpha_{\rm eff}^3 \over 4n^3}\,.
$$
Introducing the resonances and the finite width, the cross section for
the $n$th bound state becomes $\sigma_n(E_n)J_n(s,E_n)$
and $\Gamma$ continues to be twice the top weak decay width as
branching ratios to any other channel is negligible.  The effect of
the large magnitude of the top width of $1.5 $ GeV, is to smear out
the resonance structure of the bound states and yield a smooth
contribution.

The total production cross section sums the contribution of bound and
continuum states.  The resulting continuum cross sections are enhanced
over the tree level values, the effect being most pronounced just
above threshold and tapering off at very high energies. This is
demonstrated in Fig.\ 1, where the total cross section for $e^+ e^-
\rightarrow t\bar t$, calculated by including the width effects and
the QCD correction factor $K^{(f)}$ for the continuum states, is shown
as the solid curve and the tree-level cross section for the continuum
states
is shown as the
dashed curve for comparison.  The large enhancement at low relative
$t\bar t$ velocities induced by the distortion corrections in this
region is clearly visible. The cross sections remain enhanced over the
lowest order results as the energy increases.  At higher energies the
cross sections fall with the usual inverse $s$ behavior.

We also present in Fig.\ 2 the detailed threshold behavior, with the
separate contributions from the bound and the continuum states.  As
remarked earlier, the large top width smears out the resonance
structure into a smooth contribution.  These results are in conformity
to earlier investigations of the threshold region, that were spread
over wide-ranging possible top mass
values\cite{Fad90,Str91,Kwo91,Sum93}.

We have used the reported central value of 174 GeV for the top mass
\cite{Abe94}. This value is in
line with the earlier direct searches that set a lower limit $m_t >
131$ GeV/c$^2$ \cite{Aba94} and with the global fits to precision
electroweak fits that give a central value of $m_t=177$ GeV/c$^2$
\cite{Pie94}.
The accepted value of 91.17 GeV \cite{Rev92} has been used for the $Z$
mass and the weak mixing angle has been taken as $\eta =0.225$
\cite{Arr94,Rev92}.

The correction factor $K^{(f)}$ used here is similar to those used
earlier\cite{App75,Bar80,Gus88,Fad88,Fad90,Sch73}, for lighter quarks,
with small differences.  We use relativistic kinematics to interpolate
the correction factor to the high-energy region.  Our results are in
general agreement with the earlier works, but serve to quantify and
extend them using the latest values of the electroweak parameters and
top mass and are directly applicable over a wide range of energies.

\section{ Lighter heavy quarks}

  In order to demonstrate the reliability of our predictions for top
quark production, it is necessary to show that the large enhancement
of heavy quark production is consistent with experimental data.
Therefore, we extend our analysis to cover the lower energy region
spanned by experiments for heavy quark formation by $e^+ e^-$
annihilation.

  The most appropriate quantity to use for meaningful comparison with
experiment is the ratio of hadronic to leptonic (muonic) branching
ratios, commonly designated as $R$,
$$ R =
{\sigma ( e^+ e^- \rightarrow  {\rm hadrons}) \over
\sigma(e^+ e^- \rightarrow
\mu^+ \mu^-)}. $$
The numerator  takes the form
$$
\sigma(e^+ e^- \rightarrow {\rm hadrons}) =
N_c \sum_f
K^{(f)}(q_f)
\sigma_f(s),$$
where for the continuum states
$$
\sigma_f(s)=
\biggl ( {e_f \over e} \biggr )^2
{4\pi \over3 s} \alpha^2  \sqrt{ 1 - {4 m_f^2 \over s}}
\biggl (1 +  {2m_f^2 \over s} \biggr )
\theta \biggl ({ 1 - {4 m_f^2 \over s}} \biggr )   . $$
In this case the total widths being of Mev order have no perceptable
influence on flavor production cross sections. Nor do they distort the
resonance levels, which can be introduced at the appropriate
below-threshold points.

The results of the $R$ ratio calculated for the continuum states by
including the correction factor $K^{(f)}$ using
$\Lambda=\Lambda^{(3)}=338$ MeV in Eq.\ ({\ref{eq:coup})
for $c$ and $b$ production
are shown as the solid curve
in Fig.\ 3 and are compared with the experimental data \cite{Rev92}.
The $R$ ratio obtained just from the tree-level diagram without the
correction factor is also shown as the dashed curve.  The results of
Fig.\ 3 show that the inclusion of the correction factor gives a good
description of the general features of the
$R$ ratio.  The experimental $R$ ratio and the
corrected ratio $R$ are much greater than the tree-level results, near
the threshold region.  The large enhancement of heavy-quark
production is consistent with experimental data, lending support to
the predictions of Fig.\ 1 and Fig.\ 2 for $t\bar t$ production.

 The exhaustive fits to the energy dependence of $R$ beyond charm
threshold by Barnett et al\cite{Bar80} also used the interpolation
formula due to Schwinger and included the low energy dependence of the
cross section through an inverse velocity term. They used smearing
procedures for proper match between theory and experimental data above the
threshold and investigated new hypothetical particles.  They
expect their threshold results to be unreliable due to their inclusion
of the distortion effect only up to the first order in $\alpha_s$. In
the present work, we incorporate the distortion corrections to all
orders of the coupling constant on the one hand and we interpolate to
the massless quark limit on the other hand.
The simple analytical semi-empirical correction factor
$K^{(f)}$ of Eq.\ (\ref{eq:kvall}) obtained
with only a Coulomb-like potential
reproduces the gross features of $R$ over a coarse mesh of
energy, as the energy scale of Fig.\ 3 and the need for the
smearing procedure of
Ref.\ \cite{Bar80} will indicate.   In finer energy resolutions, the
experimental $R$ ratio is modulated by the presence of resonances
above the continuum threshold which necessitates the description with
a more refined potential including the additional influence of the
long-range part of $Q$-$\bar Q$ interaction, as discussed well in
Refs.\ \cite{Eic80,Zam85,Bye94}.  The agreement of our results with the
gross features of the experimental $R$ ratio adds credence to the
approximate validity of the calculated $t\bar t$ production results.

\section {Discussion.}

The importance of utilizing the largest possible cross section for top
production is self evident. In the light of possible manifestations of
new physics expected for the top era \cite{Eic94b,Lan94,Hil94} and
Standard Model violating effects
\cite{Ber94}, it is important to have a meaningful sample of top events.
It is also imperative to have a clear quantifications of Standard
Model predictions to enable clean extraction of exotic physics
signals.  Our distortion corrected cross sections provide the expected
top yield within the framework of the Standard Model and can serve as
the correct reference to probe new physics frontiers that might be
discernible beyond the top threshold.

   The large top mass introduces another feature of interest that
merits discussion. This is the coupling of the Higgs boson to the
$t\bar t$ pair, that is manifested as exchange of the scalar Higgs
boson between the members of the produced pair. Such an exchange leads
to an additional attractive potential between the $t$ and $\bar t$
that takes the Yukawa form and can be written \cite{Ina88}
$$
V_H(r)={1\over 4 \pi} \biggl ({m_t\over {\rm v}} \biggr )^2 {e^{-m_H
r}\over r}
\,,
$$
where $m_H$ and ${\rm v}$ are the mass and the vacuum expectation value of the
Higgs.  This is expected to enhance the attractive effects of gluon
exchange in the color singlet combination. However in the absence of
reliable predictions for the Higgs mass, inclusion of these effects
may be premature. Projections for Higgs effects have been carried out
by \cite{Str91}.

The possibility of toponium states bound in the pure Higgs potential
could also arise. Bounds for the exchanged particle mass, to allow
for bound states in a Yukawa potential can be estimated
\cite{Won94,Ina88}. The present lower limit of $60$ GeV for the Higgs
mass\cite{Rev94} rule out any exclusive bound states in the Higgs
potential for $t$ and $\bar t$ of the detected mass range.

Finally a comment on possible variations in the top width seems called
for.  It is well known that initial state binding and masses of the
decay products are connected to the effective decay rate of a fermion
\cite{Cha92,Cha94a}. Since in this case, the binding energy
of the decaying
top to its production partner, $\bar t$, is of the order of $0.1$ GeV
only, these effects are small and expected to be beyond the
sensitivity range of experimental access at this time. The influence
of momentum dependence on the top width and production has been
addressed recently\cite{Jez92}.

It may be pointed out that the enhancement due to the exclusive
color-singlet $t\bar t$ formation in electroweak production will not
occur in $q\bar q$ annihilation processes in $p\bar p$ and $pp$
collisions at Tevatron and LHC. Nevertheless, the influence of
distortion effects in these environments merits a discussion.  The
gluon mediated $q\bar q \rightarrow Q\bar Q$ process is suppressed by
the distortion correction \cite {Cha94} due to the repulsive
color-octet nature of the one-gluon exchange potential between $q$ and
$\bar q$ at the initial vertex and between $Q$ and $\bar Q$ in the
final vertex.  However the magnitude of this suppression is much
smaller than the enhancement for the color-singlet electroweak
channels because of the different color factors for the two cases.
The distortion effects are not as striking for QCD vector boson
mediated $q\bar q$ processes as for electroweak boson mediated ones.
However in hadronic and nuclear collisions, $q\bar q$ processes are in
any case suppressed compared to gluon induced ones. We are
investigating the distortion corrections on the gluon fusion channels
in a separate work.

 We reiterate the justification of the use of the perturbative color
potential as acting between both bound and continuum states due to the
double influence of the large virtuality of the process, corresponding
to contributing distances of order $1/m_t$ and the large top decay
rate which serves to kill the top before it can dress itself or cover
the wider range of toponia levels.  Our results, using the
experimentally reported top mass can be used directly for projections
of top flavored event rates at the proposed New Linear Collider (NLC)
facility and may help to select the most efficient colliding energies
for productive top hunting by $e^+ e^-$ annihilation.

\acknowledgments
This research was supported by the Division of Nuclear Physics, U.S.
Department of Energy under Contract No.\ DE-AC05-84OR21400 managed by
Martin Marietta Energy Systems, Inc.  One of us (LC) would like to
thank F.\ Plasil and M.\ Strayer of Oak Ridge National Laboratory for
their kind hospitality, and University Grants Commission of India for
partial support.  The authors would like to thank T.\ Barnes, W.\
Bugg, E.\ Hart, Y.\ Kamyshkov, K.\ Read, and S.\ Willenbrock for
helpful discussions.

\vfill\eject

\vfill\eject

\begin{figure}
\caption{The cross section for  $e^+ e^-
\protect\rightarrow t
\protect\bar t$.
The solid curve is the total
cross section by including the
bound and continuum states,
and the correction factor
$K^{(f)}$.  The dashed curve is obtained from the tree-level
diagram for the continuum states. \label{fig1}}
\end{figure}

\begin{figure}
\caption{The cross section for  $e^+ e^-
\protect\rightarrow t
\protect\bar t$ near the threshold region.
The solid curve is the total cross section by including the bound
states and the continuum states.  The dot-dashed curve
and the long-dashed
curve are obtained from the bound states and the continuum states
respectively.  \label{fig2}}
\end{figure}

\begin{figure}
\caption{
The ratio $R$ as a function of the
center-of-mass energy $\protect\sqrt{s}$.  The solid curve is the result
obtained by including the correction factor $K^{ (f)}$, and the
dashed curve is  from the tree-level diagram.
The data points are from the compilation of Ref.\protect\ \protect\cite{Rev92}.
\label{fig3}}
\end{figure}

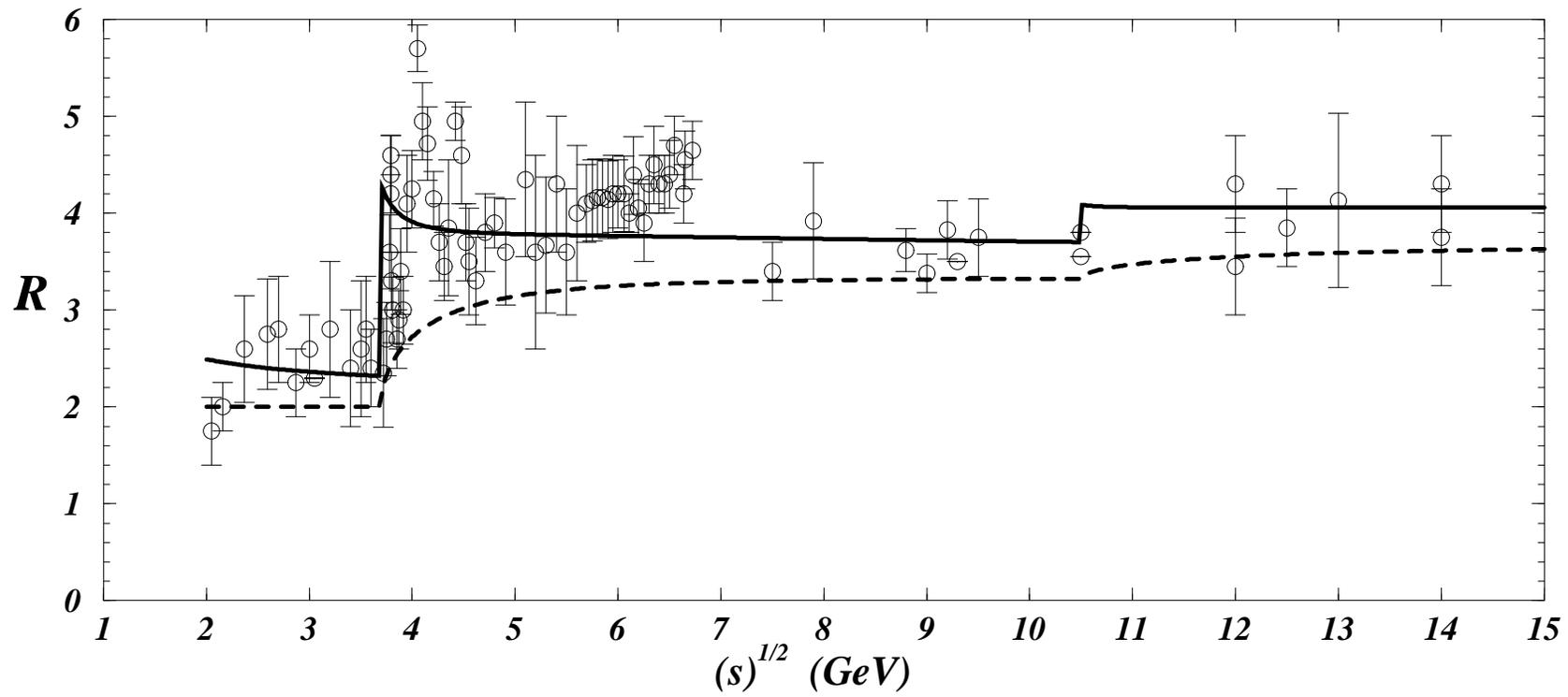
\end{document}